\documentclass[preprint,12pt]{elsarticle}

\usepackage[greek,english]{babel}
\usepackage[T1]{fontenc}
\usepackage[iso-8859-7]{inputenc}
\usepackage{epsfig}
\usepackage{xspace}
\usepackage{epstopdf}
\usepackage{verbatim}
\usepackage{axodraw4j}
\usepackage{pstricks}
\usepackage{comment}
\usepackage{bibunits}
 \usepackage{hyperref}

\biboptions{comma,square,sort&compress}

\newcounter{bla}

\journal{Comput.Phys.Commun.}

\newcommand{\eqn}[1]{Eq.(\ref{#1})}

\newcommand{\ben}{\begin{enumerate}}
\newcommand{\een}{\end{enumerate}}
\newcommand{\bit}{\begin{itemize}}
\newcommand{\eit}{\end{itemize}}
\newcommand{\bc}{\begin{center}}
\newcommand{\ec}{\end{center}}
\newcommand{\bq}{\begin{equation}}
\newcommand{\eq}{\end{equation}}
\newcommand{\bqa}{\begin{eqnarray}}
\newcommand{\eqa}{\end{eqnarray}}

\newcommand{\nl}{\nonumber\\}

\begin{document}

\def\today{October 7, 2011}
\begin{frontmatter}



\title{{\normalsize \hspace{\fill} DEMO-INP-HEPP-2011-4 \\  \hspace{\fill} WUB/11-13 \\ \hspace{\fill} TTK-11-43 \\ \hspace{\fill} IFJPAN-IV-2011-7} \\[4ex]
       HELAC-NLO
}




\author[a]{G.~Bevilacqua}
\author[a]{M.~Czakon}
\author[b,c]{M.V.~Garzelli}
\author[d]{A.~van Hameren}
\author[b]{A.~Kardos}
\author[e]{C.G.~Papadopoulos\corref{author}}
\author[f]{R.~Pittau}
\author[g]{M.~Worek}


\cortext[author] {Corresponding author.\\\textit{E-mail address:} costas.papadopoulos@cern.ch}
\address[a]{Institut f\"ur Theoretische Teilchenphysik und Kosmologie, RWTH Aachen University, D-52056 Aachen, Germany}
\address[b]{Institute of Physics, University of Debrecen, H-4010 Debrecen P.O.Box 105, Hungary}
\address[c]{University of Nova Gorica, Laboratory for Astroparticle Physics, SI-5000 Nova Gorica, Slovenia}
\address[d]{The H.\ Niewodnicza\'nski Institute of Nuclear Physics, Polisch Academy of Sciences, Radzikowskiego 152, 31-342 Cracow, Poland}
\address[e]{Institute of Nuclear Physics, NCSR Demokritos, 15310, Athens, Greece}
\address[f]{Departamento de Fisica Teorica y del Cosmos, Universidad de Granada, E-18071 Granada, Spain}
\address[g]{Fachbereich C, Bergische Universit\"at Wuppertal, D-42097 Wuppertal, Germany}
\begin{abstract}
Based on the OPP technique and the \texttt{HELAC} framework,
\texttt{HELAC-1LOOP} is a program that is capable of numerically
evaluating QCD virtual corrections to scattering amplitudes. A
detailed presentation of the algorithm is given, along with
instructions to run the code and benchmark results. The program is
part of the \texttt{HELAC-NLO} framework that  allows for a complete
evaluation of QCD NLO corrections.   

\end{abstract}


\end{frontmatter}




\begin{bibunit}[plain]
{\bf PROGRAM SUMMARY}

\begin{small}
\noindent
{\em Manuscript Title:}  \texttt{HELAC-NLO}                                     \\
{\em Authors:} G.~Bevilacqua, M.~Czakon, M.V.~Garzelli, A.~van Hameren, A.~Kardos, C.G.~Papadopoulos, R.~Pittau, M.~Worek                                               \\
{\em Program Title:}      \texttt{HELAC-1LOOP}                                      \\
{\em Journal Reference:}                                      \\
{\em Catalogue identifier:}                                   \\
{\em Licensing provisions:}    none                               \\
{\em Programming language:}       Fortran (\texttt{gfortran}\footnote{\url{http://gcc.gnu.org/fortran/}}, 
\texttt{lahey95}\footnote{\url{http://www.lahey.com}}, \texttt{ifort}\footnote{\url{http://software.intel.com}})                            \\
{\em Operating system:} Linux, Unix, Mac OS                                    
{\em Classification:}       11.1                                  \\
 {\em Nature of problem:}\\ 
  The evaluation of virtual one-loop amplitudes for multi-particle
  scattering is a long-standing problem~\cite{Ellis:2011cr}.  In
  recent years the OPP reduction technique~\cite{Ossola:2006us} opened
  the road for a fully numerical approach based on the  evaluation of
  the one-loop amplitude for well-defined values of the loop momentum.
  \\ {\em Solution method:}\\ 
  By using
  \texttt{HELAC}~\cite{Kanaki:2000ey,Papadopoulos:2000tt,Cafarella:2007pc}
  and \texttt{CutTools}~\cite{Ossola:2007ax}, \texttt{HELAC-1LOOP} is
  capable to evaluate QCD virtual
  corrections~\cite{vanHameren:2009dr}.  The one-loop $n$-particle
  amplitudes are constructed as part of the $n+2$ tree-order ones, by
  using the  basic recursive algorithm used in \texttt{HELAC}. A Les
  Houches Event (LHE) file is  produced, combining the complete
  information from tree-order and virtual one-loop contributions. In
  conjunction with real corrections, obtained with the use of
  \texttt{HELAC-DIPOLES}~\cite{Czakon:2009ss}, the full NLO
  corrections can be computed.  The program has been successfully used
  in many applications.  \\
{\em Additional comments:}\\Program obtainable from: \url{http://helac-phegas.web.cern.ch/helac-phegas}
 \\
{\em Running time:}\\
 Depending on the number of particles and generated events from seconds to days. 
   \\

\end{small}
\end{bibunit}


\section{Introduction}
\label{intro}

With the advance of LHC experiments, more precise theoretical
predictions will be indispensable. In recent years many groups have
been able to compute NLO corrections for multi-particle processes a
task thought almost impossible
before~\cite{Binoth:2010ra,Berger:2008sj,Ellis:2008qc,Berger:2009zg,Bevilacqua:2009zn,Bevilacqua:2010ve,Melnikov:2010iu,Bredenstein:2009aj,
  Bredenstein:2010rs,Bevilacqua:2010qb,Denner:2010jp,
  Bevilacqua:2011hy,Berger:2010zx,Ita:2011wn,Melia:2011dw}.  The
\textit{NLO revolution}~\cite{Salam:2011bj,Ellis:2011cr} became
possible because new reduction techniques have been proposed and
implemented, among them the so-called  OPP
technique~\cite{Ossola:2006us,Ossola:2007bb,Ossola:2007ax,Ossola:2008xq}
that allows for a fully numerical evaluation of the one-loop virtual
amplitude.

In this paper we describe one of the computational frameworks that has emerged
during the last years, namely  \texttt{HELAC-NLO}.  It
incorporates several pieces of developed software, including
\texttt{HELAC-PHEGAS}~\cite{Kanaki:2000ey,Papadopoulos:2000tt,Cafarella:2007pc},
\texttt{CutTools}~\cite{Ossola:2007ax},
\texttt{HELAC-DIPOLES}~\cite{Czakon:2009ss},
\texttt{OneLOop}~\cite{vanHameren:2010cp}, that have been already
public for some time, and \texttt{HELAC-1LOOP}, which  is presented in
this paper.  This program in its current form is capable of evaluating 
fully numerically virtual QCD corrections to scattering amplitudes
involving up to  seven particles directly attached to the loop composed by
strongly interacting particles (gluons and quarks).  In Section
\ref{helac1loop}, we will briefly describe the underlying algorithm
providing information on the structure of the code that may be useful
for potential developers. In Section \ref{run}, detailed
instructions on running the code are given. Benchmark results can be
found in Section \ref{bench}, along with a description of
\texttt{HELAC-DIPOLES} in order for the reader and potential user to have
a complete overview of the full software. Finally in Section
\ref{outlook}, we  critically  review the current level of
sophistication and the potential improvements.

\section{The \texttt{HELAC-1LOOP} algorithm}
\label{helac1loop}

The aim of the program is to numerically evaluate the virtual
contributions needed in a next-to-leading order calculation.  To this
end, the evaluation of both tree-order and one-loop amplitudes is
necessary.

The tree-order calculation algorithm is described in detail in
~\cite{Kanaki:2000ey,Papadopoulos:2000tt,Cafarella:2007pc}.  In
\texttt{HELAC-1LOOP} the first step is the construction of the
so-called tree-order skeleton, that contains all information for the
evaluation of the amplitude. The primary input to this construction is
the flavor (\texttt{ifl(1:n)}) assignment of the $n$ (\texttt{n})
external particles. A special file that is manageable by the user, also used
in {\tt HELAC}, named {\tt constants.h}, is used  to numerically fix
all physical constants needed ({\tt physics.f}).

The first action (\texttt{helac\_init} in \texttt{mastef\_new.f}) is
to enumerate (\texttt{ncc}) and define all possible color connections
(\texttt{icol(1:n,1:2)}). For a typical process  consisting of $n_q$
numbers of (outgoing)incoming (anti)quarks and $n_g$ gluons, the
number of color connections is a priori set to $(n_q+n_g)!$. Based on
these data, the program now constructs (within \texttt{pan1.f})
using a top-down approach, all currents\footnote{Notice that the
  number of currents needed is not a priori known;
  \texttt{ngues.h} contains a user-defined estimate of it.}
(\texttt{list(1:ngues,1:18,1:ncc)}) needed in a Dyson-Schwinger
recursive representation of the amplitude, using the appropriate
vertex functions (\texttt{v3} for 3-vector-boson vertex, \texttt{v4}
for 4-vector-boson vertex, \texttt{vff} for vector-boson-fermion
vertex, etc.), as shown schematically in the following figure.

\includegraphics[width=13cm]{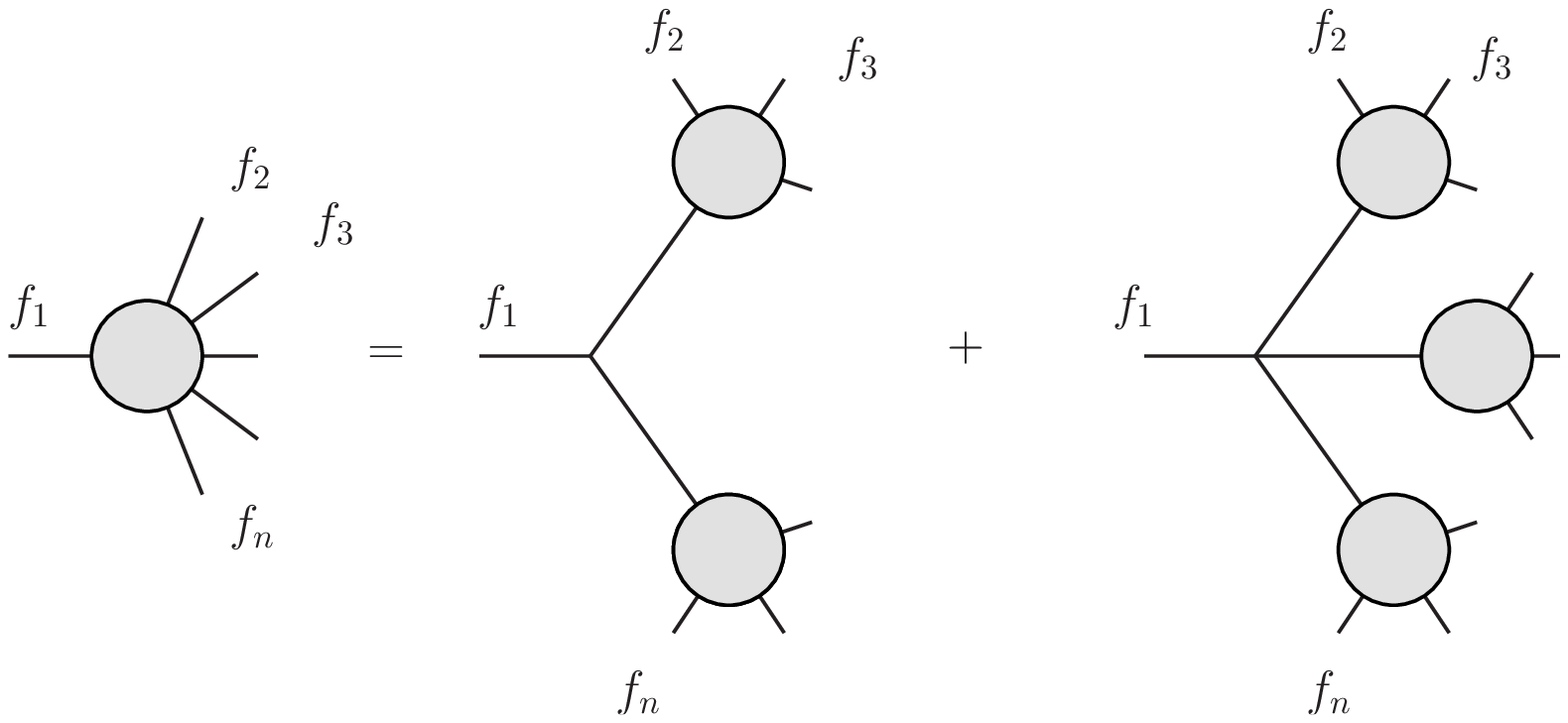}

Notice that, if in partitioning the particles within blobs in the
above diagram we keep the order of particles untouched, this is
nothing but the so-called Berends-Giele~\cite{Berends:1987me}
recursive representation for color-ordered amplitudes.  At the end of
this skeleton construction the set of all currents for all color
connections is stored in \texttt{list}. Notice that the number of
color connections  (\texttt{ncc})  can now be less than its a priori
defined value.

The one-loop $n-$particle amplitude can schematically be decomposed
in a sum over terms of the form ($ m=1,\ldots, n$)
\bqa \int \frac{\mu^{4-d} d^d \bar{q}}{(2\pi)^d}
\frac{\bar{N}(\bar{q})}{\prod_{i=0}^{m-1}
\bar{D}_i(\bar{q})} \label{def} ~, \eqa
with $d$-dimensional denominators
\bqa
 \bar{D}_{i}(\bar{q})=(\bar{q}+p_{i})^2-m_{i}^2 \, ,
\eqa
where $\bar{q}$ is the loop momentum in $d$ dimensions and
$\bar{N}(\bar{q})$ is the numerator calculated also in $d$
dimensions~\footnote{When speaking about numerator function, it
should be kept in mind that it generally contains propagator
denominators not depending on the loop momentum}. 
The sum
includes of course all terms with different loop-assignment
structure: two structures may differ either by the number
of denominators or by the different flavor and momenta appearing in
the denominators. In that
sense a closed gluon, ghost or massless quark loop, for instance,
with the same momentum flow, is considered as a different structure,
although the denominators are identical. For the highest number of
denominators each loop-assignment structure (taking into account the
flavor of the particles running in the loop) corresponds to a unique
Feynman graph, but for $m<n$ a collection of Feynman graphs with
common loop-assignment structure should be understood.

It is a well-known fact that, when the $d\to4$ limit is taken, the
amplitude can be cast into the form \bqa {\cal A}= \sum_i d_i
{\rm ~Box}_i +\sum_i c_i {\rm ~Triangle}_i +\sum_i b_i {\rm
~Bubble}_i +\sum_i a_i {\rm ~Tadpole}_i +  R \,, \label{intred}\eqa
where Box, Triangle, Bubble and Tadpole refer to the well-known
scalar one-loop functions and $R=R_1+R_2$ is the so-called rational
term.

The reduction of \eqn{def} to \eqn{intred} is the first ingredient
of any approach aiming at the calculation of virtual corrections. In
the following, we will follow the so-called \emph{reduction at the
integrand level}, developed by Ossola, Papadopoulos and
Pittau~\cite{Ossola:2006us}. The main idea is that any numerator
function can be written as

\bqa
\label{redint}
N(q) &=&
\sum_{i_0 < i_1 < i_2 < i_3}^{m-1}
\left[
          d( i_0 i_1 i_2 i_3 ) +
     \tilde{d}(q;i_0 i_1 i_2 i_3)
\right]
\prod_{i \ne i_0, i_1, i_2, i_3}^{m-1} D_{i} \nl
     &+&
\sum_{i_0 < i_1 < i_2 }^{m-1}
\left[
          c( i_0 i_1 i_2) +
     \tilde{c}(q;i_0 i_1 i_2)
\right]
\prod_{i \ne i_0, i_1, i_2}^{m-1} D_{i} \nl
     &+&
\sum_{i_0 < i_1 }^{m-1}
\left[
          b(i_0 i_1) +
     \tilde{b}(q;i_0 i_1)
\right]
\prod_{i \ne i_0, i_1}^{m-1} D_{i} \nl
     &+&
\sum_{i_0}^{m-1}
\left[
          a(i_0) +
     \tilde{a}(q;i_0)
\right]
\prod_{i \ne i_0}^{m-1} D_{i} \nl
     &+& \tilde{P}(q)
\prod_{i}^{m-1} D_{i}\,, \eqa
where now $N(q)$ and $D_{i}(q)$ are the four-dimensional versions of
$\bar{N}(\bar{q})$ and $\bar{D}_i(\bar{q})$. The coefficients $d$,
$c$, $b$ and $a$ appearing in \eqn{redint} are independent of the loop
momentum and are the same as the ones in \eqn{intred}, whereas the new
coefficients $\tilde{d}$, $\tilde{c}$, $\tilde{b}$, $\tilde{a}$ and
$\tilde{P}(q)$,  called also spurious terms, depend on the loop
momentum and they integrate to zero.

Depending on the reduction method used, the calculation of any
one-loop amplitude is placed in a very different perspective. For
instance \eqn{redint} can be solved by computing the numerator 
functions for specific values of
the loop momentum, that are solutions of equations of the form
\bqa D_i(q)=0,\,\,\, {\rm for}\,\, i=0,\ldots,M-1 \, .\eqa
It is customary to refer to these equations as quadruple ($M=4$),
triple ($M=3$), double ($M=2$) and single ($M=1$) cuts.

Calculating the numerator function for specific values of the loop
momentum opens the possibility of using \emph{tree-level amplitudes}
as building blocks. The reason is rather obvious: the numerator
function is nothing but a sum of individual Feynman graphs with the
given loop-assignment structure and as we will see in a while, it is
part of a tree amplitude with $n+2$ particles. This is by itself a
very attractive possibility, since one can use existing algorithms
and tools that perform tree-order amplitude calculations, exploiting
their automation, simplicity and speed. Indeed in what follows we will
describe how by using {\tt HELAC} we can also compute \emph{any
one-loop amplitude}.

For the one-loop amplitude, a skeleton construction is also
performed~\cite{vanHameren:2009dr}.  For a given external
configuration (\texttt{n}, \texttt{ifl}, \texttt{icol}),  the
construction of all topologically inequivalent partitions
(i.e., permutations) of the external particles into all possible number
of sets (blobs) is performed (\texttt{loop.f} and
\texttt{loop/loop\_new.f90}). One such contribution is schematically
represented in the following figure.
\bqa
 \psfig{figure=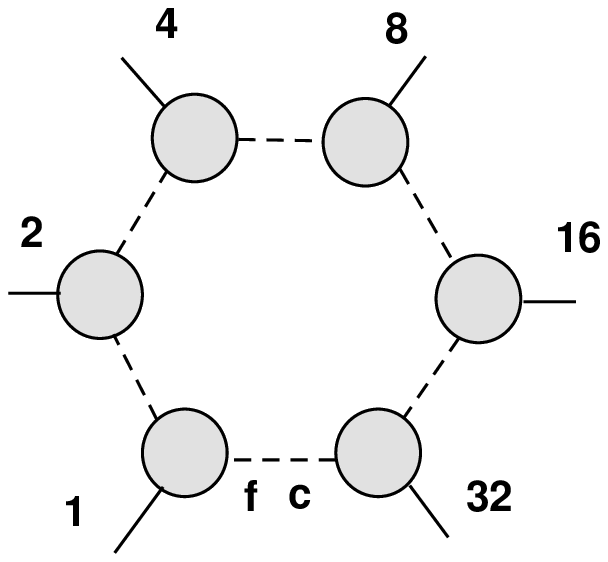,width=2truein}
\nonumber
\eqa
For those familiar with \texttt{HELAC}, the numbering of external
particles follows the binary representation used.  In the present
example, six  particles in direct contact with the loop are considered
(hexagonal topologies).  The allowed particle flavors
(\texttt{flavors.h}), and colors running in the loop are defined
(subroutines \texttt{check7}, \texttt{check6}, etc.):  the labels
\texttt{f} and \texttt{c} in the figure above refer to the possible
flavor and color of the internal loop particles.  This construction
will continue to include also pentagon topologies, tetragonal
topologies, triangle topologies and bubble topologies.  A typical
collection of possible contributions is shown in the following 
figure.
\bqa
\psfig{figure=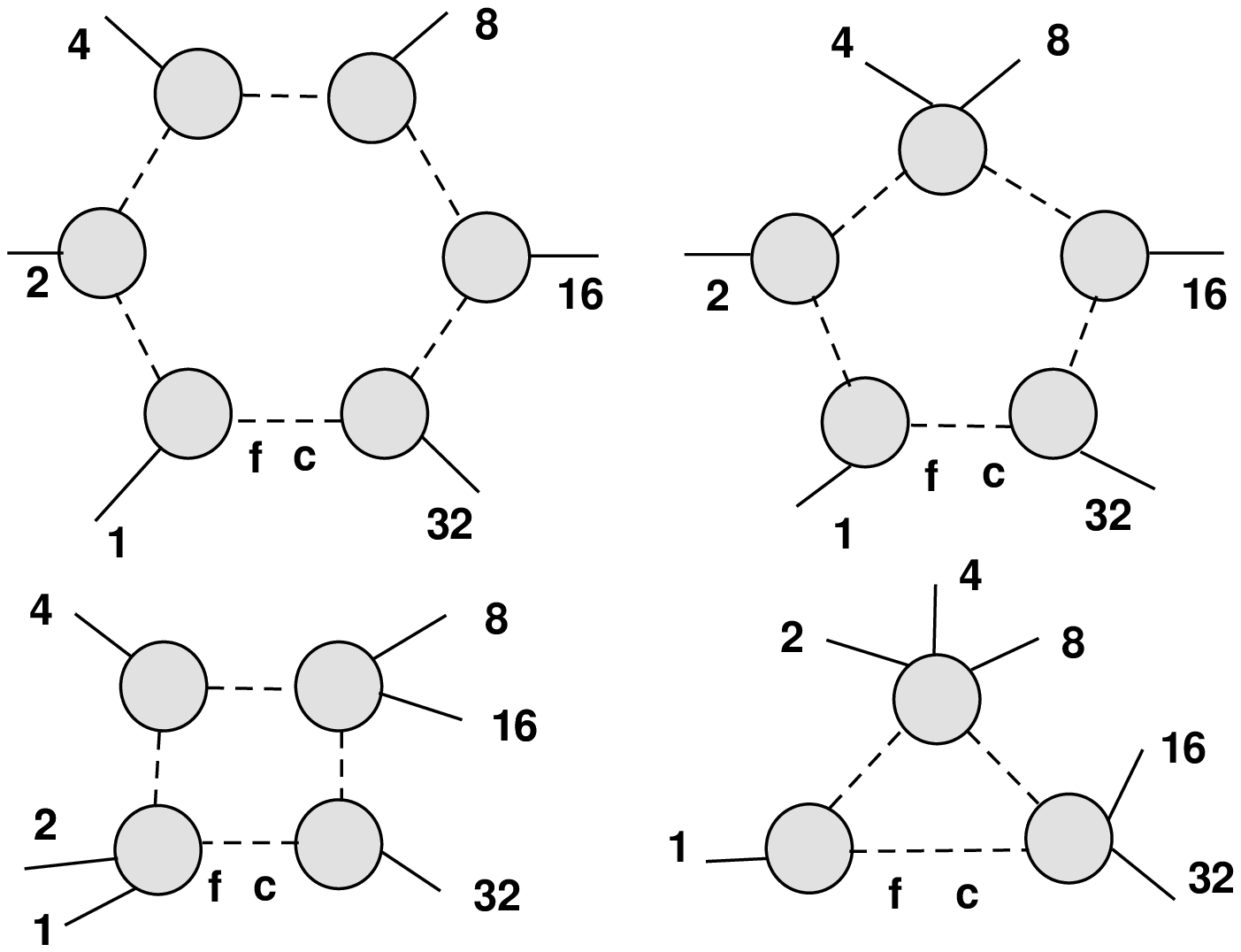,width=4truein} \nonumber 
\eqa

Concerning the loop-momentum flow in these constructions, the
convention we have chosen is that it runs counterclockwise, and the
loop propagator connecting the blob that includes the particle
number 1 and the last blob is identified as $\bar{D}_0(\bar{q})$ 
of \eqn{def}.

The selection of all the above-mentioned 
contributions is enough for the calculation of the one-loop
amplitude. To help the reader to understand the concept, the
construction we have followed is equivalent to drawing all possible
one-loop Feynman graphs, and then collecting them in subclasses that
are characterized by a common loop-assignment  structure (after
possible momentum shifts).


In practice, now, each \emph{numerator} contribution will be
calculated as part of the $n+2$ tree-order amplitude subject to the
constraint that the attached blobs will contain no propagator
depending on the loop momentum
\bqa
\psfig{figure=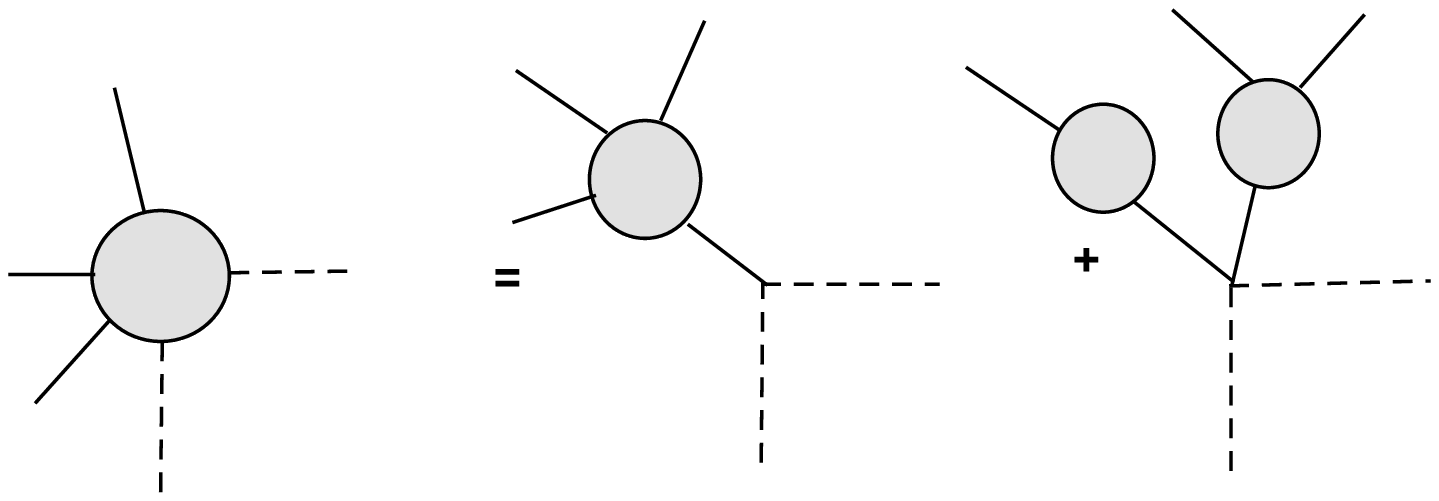,width=3truein}
\nonumber
\eqa
and no denominator will be used for the internal loop propagators.
Cutting now the line connecting the blob containing the particle
number $1$  and the last blob, it is easy to see that we have
nothing more that a part of the $n+2$ amplitude. The 'cut' particles,
with flavor \texttt{f} and color connection (\texttt{icol}) appropriately defined~\cite{vanHameren:2009dr},
will now acquire their usual numbering of
external particles in {\tt HELAC}, namely $2^{n}$ and $2^{n+1}$ (64
and 128 for $n=6$).
\bqa \psfig{figure=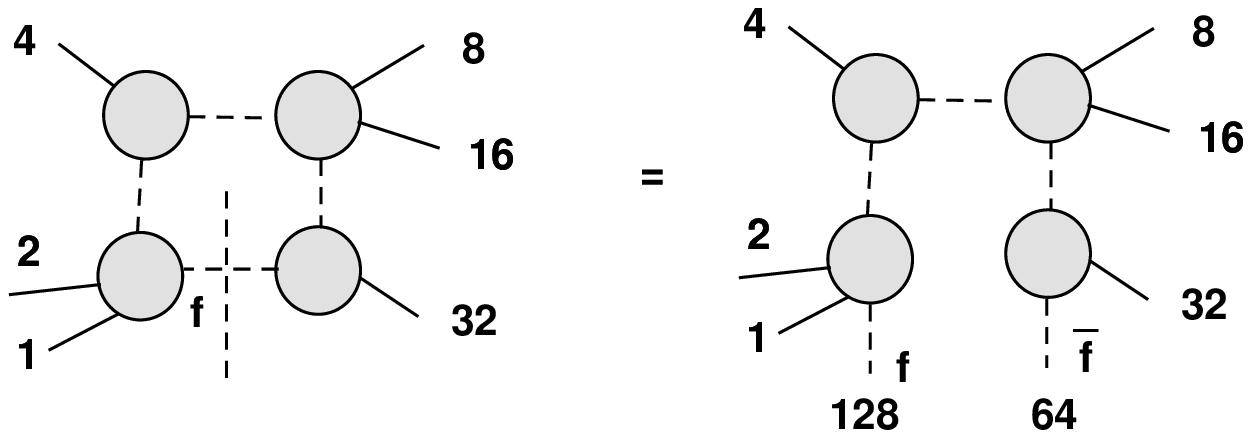,width=4truein} \nonumber
\eqa
Using the above data, the program knows how to reconstruct (subroutines
\texttt{numX\_cuttools}, \texttt{X}$=2\ldots 7$, in \texttt{loop.f})
all information needed for the calculation and store it as a sequence
of subamplitudes or currents (\texttt{listnum}), just as in the
tree-order calculation, the main difference being that for a given
color connection we have generically more than one contribution
(\texttt{nonums}) characterized by different partitions of external
particles, as well as flavors and colors running in the loop.

Rational terms are classified in two categories: the $R_1$ rational terms 
are evaluated
by \texttt{CutTools} (\texttt{cts\_xcut:rat1}); the $R_2$ rational
terms are calculated through Feynman rules as established in
~\cite{Draggiotis:2009yb,Garzelli:2009is,Garzelli:2010qm,Garzelli:2010fq}.
In \texttt{HELAC-1LOOP:physics.f} in addition to all Standard Model
couplings  defined as in \texttt{HELAC}, couplings related to the
$R_2$ rational terms are also defined.  Although we have to deal with
a tree-order construction, the skeleton is built up following a
procedure similar to that of the loop amplitude.  The reason is that
the special $R_2$ vertex has to appear only once for an arbitrary
scattering process at one loop.  Moreover these $R_2$ vertices can
join up to four particles. Therefore the skeleton construction
(\texttt{loop.f}) starts with the distribution of all external
particles  in four, three and two subsets (blobs). Then these blobs, that
represent currents or subamplitudes, are recursively
(\texttt{loop.f:mumX\_r2} with \texttt{X}=2,3,4) defined in terms of
the external particles and SM vertices, using the basic \texttt{HELAC}
algorithm.  The numerical evaluation of these contributions is
identical to that of the tree-order amplitude.  

The data collected so far are used to evaluate numerically the
amplitude in the so-called second stage.  At this moment 4-momenta
have to be supplied. In \texttt{HELAC-1LOOP} we provide three ways for
dealing with 4-momenta: the testing modes (\texttt{irambo=1},
\texttt{iint=1}) in which case the 4-momenta are generated randomly
and one can use them for a number (\texttt{nmc}) of evaluation of
the matrix element and (\texttt{irambo=0}) in which case the 4-momenta are
provided by the user through a file (\texttt{mom}); the third way
(\texttt{irambo=2}, \texttt{iint=0}), the so-called
\textit{reweighting mode},  works through the reading of a Les
Houches Event (LHE) file generated by \texttt{HELAC} itself.   

Once 4-momenta are supplied, the external wave function vectors
(four-dimensional complex vectors from \texttt{wave.f}) are computed.
Then the tree-order amplitude for each color configuration is
evaluated (\texttt{pan1.f:nextq}). The total matrix element squared is
then evaluated (\texttt{master\_new.f}) according to the standard
formula
\begin{equation}\label{fullcolorsum}
\sum_{\sigma,\sigma^\prime}A^*_{\sigma}{\cal
C}_{\sigma,\sigma^\prime}A_{\sigma^\prime} \; .
\end{equation}
The color matrix ${\cal C}$
(\texttt{rmatrix}) has been already computed and stored in the first stage.

For the one-loop amplitudes we follow the OPP reduction
(\texttt{nextq1}). To this end the \texttt{CutTools} program is
interfaced and used (\texttt{cuttools.intf.f}).  The input to
\texttt{CutTools} is the numerator function (\texttt{numerator})
evaluated at values of the loop momentum provided by \texttt{CutTools}
itself.  In order to compute the numerator function, the polarization
vectors of the two extra 'external' particles, after the one-particle
cut, are also calculated. Within the Feynman gauge  for gauge bosons,
the sum over four different (four-dimensional) polarizations that
satisfy $\sum_i e_i^\mu e_i^\nu=g^{\mu\nu}$ is performed. Ghost
particles are also included. Finally for fermions, four vectors in
spinor space, satisfying $\sum_i u^{(i)}_\alpha
u^{(i)}_\beta=(\rlap/q+m)_{\alpha \beta}$, are used.

In the \textit{reweighting mode}, the actual calculation of the
virtual  corrections is organized using a reweighting technique
~\cite{Lazopoulos:2007ix,Binoth:2008kt}.  To explain how this works,
let us start with the following equation (${\cal M}$ is the tree-order
and ${\cal L}$ the one-loop virtual matrix elements):
\begin{equation}
    \sigma^{LO+V}_{ab}=\int dx_1 dx_2 d\Phi_m f_a(x_1) f_b(x_2) \left(
|{\cal M}|^2 + {\cal M}{\cal L}^*+{\cal M}^*{\cal L} \right) \; ,
\end{equation}
which gives the sum of leading-order (LO) and virtual (V)
contributions for a scattering $ab\to m$ particles.  It can be
rewritten as
\begin{equation}
    \sigma^{LO+V}_{ab}=\int dx_1 dx_2 d\Phi_m f_a(x_1) f_b(x_2)
|{\cal M}|^2 \left(1+ \frac{{\cal M}{\cal L}^*+{\cal M}^*{\cal
L}}{|{\cal M}|^2} \right) \; .
\end{equation}
Since ${\cal L}$ is a time-consuming function one would like to
calculate it as few times as possible. To this end, a sample of
unweighted events is produced based on the tree-order distribution,
namely
\begin{equation}
   g(\vec{X})\equiv g(x_1,x_2,\Phi_m) = \frac{1}{\sigma^{LO}}
   \frac{d\sigma^{LO}_{ab}}{dx_1 dx_2 d\Phi_m} \; ,
\end{equation}
satisfying $\int d\vec{X} g(\vec{X})=1$. The sample $S$ of
unweighted events has the following property:
\begin{equation}\label{rew}
   \frac{1}{N_S}\sum_{i\in S} {\cal O}(\vec{X}_i) = \int d\vec{X}
   g(\vec{X}) O(\vec{X}) \; ,
\end{equation}
where the equality should be understood in the statistical sense,
and ${\cal O}(\vec{X})$ is any well-defined function over the
integration space. Now it is trivial to see that if
\begin{equation}
   {\cal O}(\vec{X}) =\left(1+ \frac{{\cal M}{\cal L}^*+{\cal M}^*{\cal
L}}{|{\cal M}|^2} \right) \; ,
\end{equation}
then
\begin{equation}
   \frac{1}{N_S}\sum_{i\in S} {\cal O}(\vec{X}_i)
   =\frac{\sigma^{LO+V}_{ab}}{\sigma^{LO}_{ab}} \; .
\end{equation}

In practice the sample of tree-order unweighted events includes
all information on the integration space, namely, the color
assignment, the (random) helicity configuration, 
the fractions $x_1$ and $x_2$ and the $m-$body phase space. For
future convenience, it is produced in a standard Les Houches 
format \cite{Alwall:2006yp}.
One-loop contributions are only calculated for this sample of
unweighted events, and the weight assigned to each of those events
is given by
\begin{equation}
    w=\frac{{\cal M}{\cal L}^*+{\cal M}^*{\cal L}}{|{\cal M}|^2} \; .
\end{equation}
The total virtual contribution can now be easily estimated by
\begin{equation}
    \sigma^V=\left<w\right>\sigma^B \; ,
\end{equation}
where $\sigma^B$ is the Born cross section, already included in the
LHE file. Moreover, the sample of events including the
information on $w$ can be used to produce any kinematical
distribution, according to Eq.~\ref{rew}.

Most reduction algorithms suffer from numerical instabilities usually
cau\-sed by the presence of small Gram determinants.  In order to
detect the phase space points with unstable behavior with respect to
the reduction procedure,  one can rely on several
tests~\cite{Pittau:2010tk}. In the current version of
\texttt{HELAC-1LOOP} we actually employ the so-called gauge test. Our
experience shows that the effect of numerical instabilities is more
pronounced when higher-rank tensor integrals are involved (namely high
powers of loop momentum in the numerator function). This is often
correlated with the presence of gluons in the scattering amplitude
under consideration.  In that case the gauge test can be performed, and
it is equivalent to replacing the polarization vector of an external
gluon with its momentum. Events that fail to obey the Ward identity
are separated from the initial sample. They can be treated then in
higher  (quadruple) precision. In any case, experience shows that only
for very complicated processes, like for instance $gg\to t{\bar t} gg
$, is the  effect appreciable. Care should be taken in properly
defining the numerical criterion of rejection.  For the moment we
replace the gluon polarization vector with the gluon momentum
normalized to its energy $\epsilon^\mu\to p^\mu/p^0$ and  reject
events when the computed matrix element differs from zero by an amount
greater to $10^{-9}$.   On the  other hand we should emphasize that
this procedure cannot be the final answer, since it has a limited
applicability depending on the presence of external gluons in the
scattering process under consideration.   

We should also notice that for being able to test the calculations we
provide the numerical evaluation of the infrared part (\texttt{w1\_I})
of the so-called $I$ operator.  This is achieved as described
in~\cite{vanHameren:2009dr} by computing the color-correlated matrix
(\texttt{rmatrix\_I}) and then using the tree-order amplitudes.

\section{How to run the code}
\label{run}

The code is written in \texttt{Fortran 90} and needs no additional software to run. 
We use the Fortran compiler \texttt{gfortran} as the default, but in many applications we have used also
\texttt{lahey95} and \texttt{ifort}.  
Unpacking the distributed tarball \texttt{helac1loop.tgz}
will create the directory \texttt{HELAC1L\_OFFICIAL}. In this directory you will find the 
subdirectories \texttt{examples}, \texttt{run}, \texttt{src/1LOOP}, \texttt{src/utils} and \texttt{src/TREE}.
For the convenience of the potential user, three script files, named \verb#run_testing#,  \verb#run_reweight# and  \verb#run_GC#,
can be found in the subdirectory \verb#run# that can guide the user to run the code as described in detail below.

\subsection{1LOOP}

\begin{itemize}

\item Edit the file \texttt{constants.h} and define your own physical parameters if needed (see \cite{Cafarella:2007pc} for explanation).

\item Run the script file  \texttt{scriptmake} \\
\texttt{./scriptmake n1 n2 n3 n4},
where \verb#n1# 0 means normal double precision and 1 quadruple (if supported by fortran version),
\verb#n2# 0 for compiling everything from scratch and 1 only the latest files,  \verb#n3# is empty by default
and can be used to pass additional flags to fortran commands (see the \texttt{scriptmake} file),  
and finally  \verb#n4# either empty or \verb#GC#, which means gauge check mode. 
For instance, the command
\begin{verbatim} 
./scriptmake 0 0 "" "GC"
\end{verbatim} 
will produce the \texttt{main\_onep\_dpGC.exe} executable file, for use in gauge check,
where 
\begin{verbatim} 
./scriptmake 0 0
\end{verbatim} 
will produce the \texttt{main\_onep\_dp.exe} executable file.

\item Edit the input file \texttt{input} and define appropriately the input parameters:

\begin{itemize}
\item \texttt{iint} 0 for reweighting mode, 1 for all others
\item \texttt{ibv} 0 for full summation over colors, 1 for Monte Carlo (MC) over colors
\item \texttt{iverbose} 0...3 different levels of verbosity
\item \texttt{repeat} 1 only skeleton construction, 2 only amplitude evaluation (assuming the skeleton is present), 0 for one shot 
\item \texttt{iranhel} 0 for full summation over helicities, 1 for MC over helicities 
\item \texttt{n} number of particles 
\item \texttt{flavors} flavor of particles according to the \texttt{HELAC} list
\item \texttt{iflag} 0 for internal definition of helicities, 1 for user providing helicities
\item \texttt{ihiggs} 0 if no Higgs is included, 1 if Higgs is included if allowed
\item \texttt{loopi} T if loop amplitude is calculated, F if only tree is calculated 
\item \texttt{onlyqcd} T if only QCD couplings are allowed, F if also EW couplings are allowed
\item \texttt{withqcd} T if QCD is included, F elsewhere
\item \texttt{irambo} 0 if momenta are provided by the user (through \texttt{mom} file), 1 if are generated randomly by \texttt{RAMBO}, 2 if provided by the
\texttt{.lhe} LHE file generated by \texttt{HELAC}
\item \texttt{e} energy in GeV
\item \texttt{nmc} number of phase-space points to be evaluated (for \texttt{irambo} 0 or 1)
\item \texttt{mom} file with the 4-momenta in the format $E$, $p_x$, $p_y$, $p_z$, $m$
\item \texttt{momout} output file for user provided momenta
\item \texttt{muscale} the renormalization scale 
\end{itemize}

\end{itemize}

\subsection{TREE}

In this directory, you can generate the LHE file, for later use in the reweighting procedure. The generation
is identical to the standard \texttt{HELAC-PHEGAS} procedure. So we refer the potential user to \cite{Cafarella:2007pc}. 
Some new elements with respect to the standard treatment have been added, without affecting the generation procedure.
\begin{itemize}
\item In this version a color MC is used according to \cite{Bevilacqua:2009zn}. There is a new keyword \texttt{color\_flag}, which is 0 for a full color summation
and 1 for an MC over colors. 
\item An interface to phase-space generator \texttt{KALEU}~\cite{vanHameren:2010gg} is also present; the user can set the \texttt{phasespace\_flag} to 0 for \texttt{PHEGAS}  or 1
for \texttt{KALEU}.
\item Finally, the \texttt{oneloop\_rewgt} keyword has to be set to true (T) in order for the generated LHE file to be used for the reweighting procedure described above.
\end{itemize}
 
 The tree-order generation will result in an LHE file, \texttt{sampleG0G0TqTa.lhe}, for instance for the process $gg\to t\bar{t}$. 
 
\section{Results and Benchmarks}
\label{bench}
In this section we will give characteristic examples of running the code. The tarball file also contains the \texttt{examples} directory
where these results are stored. 

\subsection{The testing mode}

Let us choose a relatively simple example $u\bar{d}\to W^+ + n g$ with $n=1,2,3$. We start with simplest case: $n=1$.
The input file looks like the following:
\begin{verbatim}
1                      ! iint
0                      ! ibv
3                      ! iverbose
1 0                    ! repeat,iranhel
4                      ! n
3 -4 33 35             ! flavours
0 0                    ! iflag,ihiggs 
t                      ! loopi
f t 0                  ! onlyqcd,withqcd,irun
1 500 1	"mom" "momout"    ! irambo,energy,nmc,momfile,momout
170.9                  ! muscale
\end{verbatim}
The output file (the reader can find it as \texttt{uDWg\_gen} in the official distribution \texttt{examples}) incorporates information on how the 
different currents are formatted. The main result of this run is encoded in the files 
\begin{itemize}
\item \texttt{tree\_UqDaW+G0.in}, including all the information needed by \texttt{HELAC-1LOOP} to evaluate  the amplitude numerically, 
and
\item \texttt{treeli\_UqUaW+G0.in}, with a brief description of the virtual amplitude generation.
\end{itemize}

Now, by changing the \texttt{repeat} keyword value from 1 to 2, we have the numerical evaluation of the amplitude in \texttt{uDWg\_out}. 
Most of the content of the output file is self-explanatory. 
We focus here on its major aspects.
After the printout of physical constant the user (if $\mathtt{iverbose}\ge 1$) will see the following: 
\begin{verbatim}
 UqDaW+G0
\end{verbatim}
which is a line printing out the process under consideration $u\bar{d}\to W^+ g$.
There then follows a series of lines such as 
{\footnotesize
\begin{verbatim}
 INFO =============================================
 INFO COLOR           1 out of           2
INFO   2   6  -3   5   1   1   4  34   3   2  -4   2   0   0   0   1   1   0
INFO   2  10  -4   6   1   1   8  35   4   2  -4   2   0   0   0   1   1   3
INFO   2  10  -4   6   0   1   8  35   4   2  -4   2   0   0   0   2   1   3
INFO   2  14  -3   7   1   2   4  34   3  10  -4   6   0   0   0   1   1   0
INFO   2  14  -3   7   2   2   8  35   4   6  -3   5   0   0   0   1   1   3
INFO   2  14  -3   7   0   2   8  35   4   6  -3   5   0   0   0   2   1   3
\end{verbatim}
}
where each line represents a subamplitude needed in the construction of the tree-order amplitudes corresponding to the  
\verb#1 out of           2# color connection configurations existing for this process.
Then,  after exhausting all tree-order information, you start seeing 
{\footnotesize
\begin{verbatim}
 LOOP T
 INFO =============================================
 INFO COLOR           1 out of           2
 INFO number of nums          33
 INFO NUM           1  of           33           6
INFO   3  24   3   7   1   1   8  35   4  16   3   5   0   0   0   1   1   3
INFO   3  24   3   7   0   1   8  35   4  16   3   5   0   0   0   2   1   3
INFO   3  28   4   8   1   1   4  34   3  24   3   7   0   0   0   1   1   0
INFO   1  30  35   9   1   1   2  -4   2  28   4   8   0   0   0   0   1   2
INFO   2  62  -3  10   1   1  30  35   9  32  -3   6   0   0   0   1   1   2
INFO   2  62  -3  10   0   1  30  35   9  32  -3   6   0   0   0   2   1   2
INFO   4   8   4   2   1   3   3   4  35   0   0   0   0   0   0   0   3   1
INFOYY   1
\end{verbatim}
}
again for the first color connection. As it is printed there are now 33 (\verb#number of nums          33#) contributions,
and the first one is nothing but the following box graph:
\begin{center}
  \begin{picture}(240,240) (35,-43)
    \SetWidth{1.0}
    \SetColor{Black}
    \Line[arrow,arrowpos=0.5,arrowlength=5,arrowwidth=2,arrowinset=0.2](64,-32)(96,32)
    \Line[arrow,arrowpos=0.5,arrowlength=5,arrowwidth=2,arrowinset=0.2](96,32)(176,32)
    \Line[arrow,arrowpos=0.5,arrowlength=5,arrowwidth=2,arrowinset=0.2](176,32)(176,112)
    \Line[arrow,arrowpos=0.5,arrowlength=5,arrowwidth=2,arrowinset=0.2](176,112)(96,112)
    \Line[arrow,arrowpos=0.5,arrowlength=5,arrowwidth=2,arrowinset=0.2](96,112)(64,176)
    \Gluon(176,32)(224,-16){7.5}{3}
    \Photon(176,112)(224,160){7.5}{3}
    \Gluon(96,112)(96,32){7.5}{4}
    \Text(32,-48)[lb]{\Large{\Black{$u(1)$}}}
    \Text(240,-32)[lb]{\Large{\Black{$g(8)$}}}
    \Text(240,160)[lb]{\Large{\Black{$W^+(4)$}}}
    \Text(32,176)[lb]{\Large{\Black{$\bar{d}(2)$}}}
  \end{picture}
\end{center}
The relevant topological information can easily be read off from the line
{\footnotesize
\begin{verbatim}
INFO   4   8   4   2   1   3   3   4  35   0   0   0   0   0   0   0   3   1
\end{verbatim}
}
the first number after the keyword \verb#INFO#, namely \verb#4#, being the number of loop propagators, 
then \verb#8   4   2   1# the ordering of blobs and finally \verb#3   3   4  35# ($u u d g$) the corresponding flavor assignment.
 
Out of the 33 contributions, the last six refer to $R_2$ tree order like ones including a unique special vertex each. In the following example,
a special $R_2$ $u\bar{d}W$ vertex
{\footnotesize
\begin{verbatim}
 INFO NUM          29  of           33           3
INFO  25   6  -3   5   1   1   4  34   3   2  -4   2   0   0   0   1   1   0
INFO   2  14  -3   6   1   1   8  35   4   6  -3   5   0   0   0   1   1   3
INFO   2  14  -3   6   0   1   8  35   4   6  -3   5   0   0   0   2   1   3
INFO   0   0   0   0   0   0   0   0   0   0   0   0   0   0   0   0   0   1
INFOYY   1
\end{verbatim}
}
is used as shown also schematically in the following Feynman graph:
\begin{center}
\fcolorbox{white}{white}{
  \begin{picture}(192,228) (35,-79)
    \SetWidth{1.0}
    \SetColor{Black}
    \Line[arrow,arrowpos=0.5,arrowlength=5,arrowwidth=2,arrowinset=0.2](48,-76)(112,-12)
    \Gluon(112,-12)(176,-76){7.5}{5}
    \Line[arrow,arrowpos=0.5,arrowlength=5,arrowwidth=2,arrowinset=0.2](112,-12)(112,84)
    \Photon(112,84)(176,152){7.5}{4}
    \Line[arrow,arrowpos=0.5,arrowlength=5,arrowwidth=2,arrowinset=0.2](112,84)(48,148)
    \Text(32,-28)[lb]{\Large{\Black{$u(1)$}}}
    \Text(192,-12)[lb]{\Large{\Black{$g(8)$}}}
    \Text(192,84)[lb]{\Large{\Black{$W^+(4)$}}}
    \Text(32,100)[lb]{\Large{\Black{$\bar{d}(2)$}}}
    \Vertex(112,84){16}
  \end{picture}
}
\end{center}

At the end of the file, after momenta have been generated and printed,
you find the most important result which is the numerical values of matrix elements:
{\footnotesize
\begin{verbatim}
 HELICITY CFGS           6
 total amplitude squared LO   =   0.73261294162118751
 total amplitude squared U0 =   4.88872424257442914E-002
 total amplitude squared T0 =   4.88872424257442914E-002
 ratio =   6.67299738352457872E-002
 total amplitude squared U1 =   2.38831983483179001E-002
 total amplitude squared T1 =  -2.26466317876591848E-002
 total amplitude squared T2 =  -6.87832271575313514E-002
 total amplitude squared I1 =  -2.26466317876590842E-002
 total amplitude squared I2 =  -6.87832271575313653E-002
 total amplitude squared R0 =    0.0000000000000000
 total amplitude squared R1 =  -4.65298301359770849E-002
\end{verbatim}
}
In this case,  there are six non-zero helicity configurations, \verb#LO# is the leading order $|{\cal M}|^2$
properly summed and averaged, \verb#U0# is the finite part of the one-loop amplitude 
without wave function and coupling constant renormalization
(but with the mass renormalization contribution
 if present), \verb#T0# the renormalized one, and so on for \verb#U1#, 
\verb#T1# for the $\frac{1}{\epsilon}$ and \verb#U2#, \verb#T2# for the $\frac{1}{\epsilon^2}$ terms. \verb#I1# and \verb#I2# are the
corresponding poles predicted by the $I$ operator, and the level of agreement between \verb#T2# and \verb#I2# as well as between \verb#T1# and \verb#I1# reflects
the precision achieved.

In the website of \verb#HELAC-NLO#, results can also be found for $u\bar{d}\to W^+ g g$ and $u\bar{d}\to W^+ g g g $.

\subsection{Reweighting mode}

The distributed version contains reproducible results for the process $gg \to t \bar{t}$. 
To obtain those results you have to first generate the LHE file 
in the \texttt{TREE} subdirectory by executing the command
\begin{verbatim}
./run.sh user.inp myenv
\end{verbatim}
as in the usual \texttt{HELAC-PHEGAS}. In general, the user has to edit the \verb#user.inp# input file and define the corresponding input parameters.

Copy the produced sample file \texttt{sampleG0G0TqTa.lhe} in the same directory where the \texttt{main\_onep\_dpGC.exe} is used in case
you have to run a gauge test.
To this end the set-up of the input parameters has to be as follows:
\verb#iint#$=0$, \verb#ibv#$=1$ and \verb#irambo#$=2$.
When now the executable is running two sample LHE files will be generated
\texttt{sampleG0G0TqTa\-\_GAUGECHECK\-\_FAILED.lhe} 
and \texttt{sampleG0G0TqTa\-\_GAUGECHECK\-\_PASSED.lhe}, containing the events that have failed and passed
the test respectively. 

Then rename this last file to \texttt{sampleG0G0TqTa.lhe} and run 
\begin{verbatim}
./main_onep_dp.exe < input ><<your output>>
\end{verbatim}
The final result is the LHE file \texttt{sampleG0G0TqTa\_WEIGHTED.lhe}, which can now be used
in conjunction with \texttt{HELAC-DIPOLES} to get the full NLO corrections. 

The events, if any, in
\texttt{sampleG0G0TqTa\_GAUGECHECK\_FAILED.lhe}, \\ 
can 
be reprocessed in higher numerical precision\footnote{Such a possibility can be 
realized with \texttt{lahey95} and \texttt{ifort} that incorporate at the compilation level the option of quadruple precision. 
See also the script file \texttt{scriptmake} for more
details.}:
\begin{verbatim}
./main_onep_qp.exe < input ><<your output>>
\end{verbatim}
The resulting file is again \texttt{sampleG0G0TqTa\_WEIGHTED.lhe}, which can now be combined\footnote{In the subdirectory \texttt{src/utils/COMBINE}, useful
tools for combining LHE files are
available.}  with the one produced with the passed events. Care of course has to be paid by the user in properly managing and renaming the emerging files.

\subsection{HELAC-DIPOLES}
\label{dipoles}

In this subsection we will briefly review the use of \texttt{HELAC-DIPOLES} in order for the user to be able to have a 
more complete overview of the full software. For a detailed description please refer to~\cite{Czakon:2009ss}.

Unpack the tarball  \texttt{dipoles.tgz} inside the main \texttt{HELAC} directory and copy
\texttt{alphas\_std.h} from the \texttt{dipoles} directory to \texttt{HELAC}. This sets up
two-loop running for strong coupling constant consistently with the PDFs used by \texttt{HELAC-DIPOLES}.

\begin{itemize}
\item Configure  \texttt{HELAC-PHEGAS} 
\begin{itemize}
\item Edit file \verb#myenv# to set up the libraries and compilers.
\item Edit \verb#user.inp#, to specify the input physical parameters.
\item Edit \verb#getqcdscale.h#, if a running coupling constant is needed.
\item Edit \verb#nh.h#, to specify the number of histograms that will be generated.
\end{itemize}
Apart from \verb#myenv#, for all other files described so far the user may use their default configuration. To run 
the program use the command
\begin{verbatim}
   ./run.sh user.inp myenv
\end{verbatim}
This procedure will set up several parameters needed by \texttt{HELAC-DIPOLES} later on, among others the cuts to be used 
by the integrated dipoles (to be run with \verb#make run_I# and 
\verb#make run_KP#, see below).
\item Configure \texttt{HELAC-DIPOLES}
\begin{itemize}
\item Edit \verb#dipoles.input#, to specify the process and various optimization
   parameters just as in the original \texttt{HELAC}.
\item Edit \verb#dipoles.conf#, to set specific parameters for the calculation of
   the dipoles, as follows.
\begin{itemize}
\item \verb#onlyreal#: if set to true, only real corrections for a given process
	   will be calculated. The cuts are then specified in \verb#cuts.h#.
	   The result must coincide with the one of the original
	   \texttt{HELAC} with the same input parameters. This option is
	   included for testing purposes.

\item \verb#onlylast#: if set to true, only those dipoles will be included, which
	   contain the last particle (for correctness it must be a
	   parton). This is useful for some processes, where it is
	   clear that only the last particle can be soft/collinear,
	   and the bookkeeping remains simple to obtain the full
	   result at NLO.

\item \verb#onlydiv#: if set to true, only divergent dipoles will be
	  included. Non-divergent dipoles correspond to a pair of
	  massive quarks in the final state. They are only useful to
	  get rid of large Sudakov logarithms, but are not essential
	  for the finiteness of the real radiation contribution.

\item \verb#hybrid#: if set to true, non-parton polarizations will be summed over
       	 by a continuous Monte Carlo integration over a phase parameter.

\item \verb#signmode#: defines how positive and negative contributions are to be
	   treated: 0 - the result is left unchanged, whether positive or
	       negative,
	   1 - only positive numbers, a negative result is set to zero,
	   2 - only positive numbers, but the sign of a negative
	       result is changed, and positive results are set to zero

\item \verb#sumtype#: in the first phase (preferably for phase-space
	  optimization), the summation over helicities of the partons
	  can be performed in three different ways:
	  0 - exact fast summation (independently for real radiation
	      and dipoles),
	  1 - exact slow summation (for a given helicity configuration
	      both real radiation and the dipole sum will be
	      calculated),
	  2 - flat Monte Carlo summation over all non-vanishing
	      helicity configurations. In practice, it is recommended to use the last option.

\item \verb#nsumpol#: number of accepted points to be summed over helicity with
	  the method specified by \verb#sumtype#. The counting starts after
	  phase space optimization is finished.

\item \verb#noptpol#: number of accepted points to be used for helicity sampling
	  optimization. During helicity sampling optimization, slow
	  summation over helicity configurations (in the sense defined
	  in the description of \verb#sumtype#) is performed. It is therefore
	  recommended to keep this number relatively small (of the
	  order of a few hundred to a thousand).

\item \verb#nuptpol#: number of accepted points after which an updated of the
	  helicity sampling weights is performed. This number should
	  be rather large for best results (at least an order of
	  magnitude larger than \verb#noptpol#).

\item \verb#alphaMinCut#: lowest value of \verb#alphaMin#, below which a point will be
	      rejected altogether, because of the risk of numerical
	      instabilities. For the exact definition of \verb#alpha#, see~\cite{Czakon:2009ss}.

\item \verb#alphaMaxII, alphaMaxIF, alphaMaxFI, alphaMaxFF, kappa#: parameters of
	    the dipoles (see~\cite{Czakon:2009ss}).
\end{itemize}

\item \verb#jetfunctions.f#, for non-trivial jet functions, although most
   work should be performed on FJmpo (jet function for real radiation).

\item \verb#cuts.h#, to specify general cuts to be used by the jet functions.

\item \verb#histograms.f#, to define histograms.

\item \verb#seed.input#, to change the random number seed, which is useful for
      trivial parallelization runs (this should be an integer).
\end{itemize}
Again one can use the existing default configuration. To run the program
\begin{verbatim}
export FC="<<your fortran>>"
make
make run
\end{verbatim}
This will compile the dipole-subtracted version and run it.
If this is the first run for a given process, the calculation
will be stopped,  and you should run
\begin{verbatim}
make trees
make run
\end{verbatim}
The former will generate and store in the subdirectory \texttt{helac\_trees} all skeleton files for the different subprocesses needed
by the dipole subtraction.  
\item
For the $I$ operator, the corresponding input and configuration files are
\verb#dipoles_I.input# and \verb#dipoles_I.conf#. The parameters defined in the configuration file are the renormalization scale, 
the number of light $N_f$ and heavy $N_F$ quark flavors, used in the definition of $I$ operator~\cite{Catani:2002hc,Czakon:2009ss}, the 
$\alpha$ parameter that controls the integration of dipole functions over the available phase-space and the parameter $\kappa$ defined in~\cite{Catani:2002hc}.
To run
\begin{verbatim}
make run_I
\end{verbatim}

\item
For the $K+P$ operator the corresponding input and configuration files are
\verb#dipoles_KP.input# and \verb#dipoles_KP.conf#. In addition to the common parameters described so far for the $I$ operator, the configuration file 
contains also the definition of the factorization scale and the number and flavors of initial state partons
to be taken into account. 
To run
\begin{verbatim}
make run_KP
\end{verbatim}

\end{itemize}

The result of the run consists of several files. A typical output will have at the end the following information 
{\footnotesize 
\begin{verbatim}
 out of      1000000        1000001  points have been used
 and        340079  points resulted to =/= 0 weight
 whereas        659922  points to 0 weight
  estimator x:    0.381485D-16
  estimator y:    0.143375D-32
  estimator z:    0.205540D-65
  average estimate :   0.381485D-16
                +\-    0.378649D-16
  variance estimate:   0.143375D-32
                +\-    0.143367D-32
  be aware that the error estimate may be bad!
  estimator x:    0.835522D-04
  estimator y:    0.156118D-12
  estimator z:    0.930801D-29
  average estimate :   0.835522D-04
                +\-    0.395118D-06
  variance estimate:   0.156118D-12
                +\-    0.305090D-14
 total XS -8.35522401509068940E-005  3.95117621007448708E-007
 lwri: points have used   0.0000000000000000
        2212        2212   7000.0000000000000        7000.0000000000000                3           1
 % error:   99.256586045324298
 % error:  0.47289889570121785
\end{verbatim}
}
which states that from a run of 1 million points, 340,079 have been used after cuts and the positive part of the cross section
is \verb#0.381485D-16# with a statistical uncertainty \verb#+\-    0.378649D-16# whereas the negative part is 
\verb#0.835522D-04#, \verb#+\-    0.395118D-06# (in nanobarns). See also \cite{Cafarella:2007pc} for more explanation. 
Also files named \verb#hi_file#, \verb#hi_file_I#, \verb#hi_file_KP# will be generated with all data needed for the histograms.  

\section{Outlook}
\label{outlook}

The progress in NLO calculations seen over the last years has made the development of an automatized computational framework a realistic 
task~\cite{Ossola:2007ax,Gleisberg:2007md,Berger:2008sj,vanHameren:2009dr,Czakon:2009ss,Berger:2009ba,Frederix:2009yq,Hirschi:2011pa}.
As in the case of tree-order generation, NLO programs will be able to generate LHE files ready for use in physics analyses.
Nevertheless there are several open issues to be addressed in the near future.
\begin{itemize}
\item For the moment, LHE files can be generated for tree-order plus virtual corrections. For real corrections usually one has to deal with 
so-called "weighted" events, namely a large collection of phase-space integration points. Taking into account also that 
the Monte Carlo convergence of the real corrections, especially that of the subtracted real emission part, is quite slow a solution will be 
very welcome. Within \texttt{HELAC-NLO} we plan to further investigate the possibility of using alternative phase-space algorithms, sampling
over colors and other subtraction methods in order to achieve a significant improvement in the overall efficiency.
\item The interface to parton shower~\cite{Frixione:2002ik,Nason:2004rx,Frixione:2007vw,Alioli:2010xd} 
programs is also an important issue at the phenomenological level. There are already several
steps taken towards this direction. 
So far, \texttt{HELAC-1LOOP} has been interfaced to the \texttt{POWHEG-BOX} framework
at the purpose of studying specific processes like $pp \to ttH$~\cite{Garzelli:2011vp}
and $pp \to ttj $~\cite{Kardos:2011qa}, including NLO QCD corrections matched to a
parton shower evolution, followed by hadronization and hadron decay, up to final predictions at the hadron level to be compared to LHC and Tevatron data.
We plan to further investigate this issue with the aim of integrating and automating the full procedure.
\item To address the incorporation of the full set of Electroweak corrections as well as theories beyond the SM, a reimplementation of the \texttt{HELAC}
algorithm is desirable. In such a process, we aim also to include several straightforward improvements in constraining the redundancy of the actual
computation, resulting in a significant reduction in computing time and resources. 
\end{itemize}
 
Finally it will be very interesting to advance beyond one loop. Both the OPP reduction method and the recursive approach to scattering amplitude
computation may open the road to highly efficient calculations at the two-loop level~\cite{Mastrolia:2011pr}.
\\[12pt]

{\bf Acknowledgments:} It is our pleasure to acknowledge very fruitful discussions with Z.~Bern, D.~Kosower, L.~Dixon, H.~Ita, D.~Maitre, 
Z.~Kunszt, A.~Denner, C.~Anastasiou, 
G.~Ossola, P.~Mastrolia, W.~Giele, R.~Frederix, A.~Lazopoulos, P.~Draggiotis and I.~Malamos. 
We also acknowledge partial support from FALCON FPA 2008-02984, HEPTOOLS MRTN-CT-2006-035505, LHCPhenoNet PITN-GA-2010-264564. 
M.C was supported by the Heisenberg and by the Gottfried Wilhelm Leibniz
Programmes of the Deutsche Forschungsgemeinschaft. M.W acknowledges  support by the 
Initiative and Networking Fund of the Helmholtz Association, contract HA-101 (Physics at the Terascale).
M.V.G acknowledges the University of Gra\-na\-da, Departamento de Fisica Teorica y del Cosmos, and the NCSR Demokritos, Institute of Nuclear 
Physics, for hospitality.
C.G.P and R.P would like to thank the KITP at UCSB for the kind hospitality offered during the completion of this paper.
A. K. was supported by the T\'AMOP 4.2.1./B-09/1/KONV-2010-0007 and the T\'AMOP 4.2.2/B-10/1-2010-0024  projects.
G.B. acknowledges support by the DFG Sonderforschungsbereich/Transregio 9 Computergest\"utzte Theoretische Teilchenphysik.
\\


{\bf\large References}









\end{document}